# Excellent HER and OER Catalyzing Performance of Se-vacancies in Defects-engineering PtSe$_2$: From Simulation to Experiment

Yuan CHANG, Panlong Zhai, Jungang Hou, Jijun Zhao, Junfeng Gao*

*Key laboratory of Material Modification by Laser, Ion and Electron Beams, School of Physics, Dalian University of Technology, Dalian 116024, P. R. China.*

*State Key Laboratory of Fine Chemicals, School of Chemical Engineering, Dalian University of Technology, Dalian 116024, P. R. China.*

E-mail: gaojf@dlut.edu.cn

**ABSTRACT:** Facing with grave climate change and enormous energy demand, catalyzer gets more and more important due to its significant effect on reducing fossil fuels consumption. Hydrogen evolution reaction (HER) and oxygen evolution reaction (OER) by water splitting are feasible ways to produce clean sustainable energy. Here we systematically explored atomic structures and related STM images of Se defects in PtSe$_2$. The equilibrium fractions of vacancies under variable conditions were detailly predicted. Besides, we found the vacancies are highly kinetic stable, without recovering or aggregation. The Se vacancies in PtSe$_2$ can dramatically enhance the HER performance, comparing with, even better than Pt(111). Beyond, we firstly revealed that PtSe$_2$ monolayer with Se vacancies is also a good OER catalyst. The excellent bipolar catalysis of Se vacancies were further confirmed by experimental measurements. We produced defective PtSe$_2$ by direct selenization of Pt foil at 773 K using a CVD process. Then we observed the HER and OER performance of defective PtSe$_2$ is much highly efficient than Pt foils by a series of measurements. Our work with compelling theoretical and experimental studies indicates PtSe$_2$ with Se defects is an ideal bipolar candidate for HER and OER.

# 1. Introduction

With the increasing enormous demand for energy and severe environmental pressure, searching efficient and clean sustainable energy to substitute traditional fossil fuels is urgent. Hydrogen evolution reaction (HER) and oxygen evolution reaction (OER) by electrochemical water splitting are feasible ways, which have been intensively explored.[1-5] The key for HER and OER is selecting highly efficient catalyst. Two-dimensional (2D) materials, possessing intrinsic very large surface, are ideal catalytic platform.[6]

Surface defects and modification can dramatically enhance the catalyst of 2D mateirals.[7-21] However, this requires insightful understanding of the structures, stability, properties and catalytic mechanism for defects in 2D materials. One superior method is ab initio simulations, which are high reliable in prediction of structure and stability of 2D materials, understanding the 2D growth mechanism, revealing defects and related electronic and catalysis properties. For instance, two-dimensional boron polymorphs were first probed theoretically and synthesized successfully on Ag(111) substrate.[22,23] There are indispensable contribution of theoretical predictions in the exploration of topological insulator $Sb_2Te_3$[24,25] and twisted bilayer graphene[26-29] as well.

Elementary platinum is a very good catalyst for HER,[30] however, it is too expensive. Only top surface of Pt slab serves as effective catalyst, while inner Pt atoms are waste. Besides, Pt surface cannot be used for OER. 2D atomic thin $PtSe_2$ by direct selenization of Pt surfaces,[31] reduces the amount of Pt and can provide catalyst from both sides. Although 2D $PtSe_2$ exhibits high electron mobility,[32,33] unique Dirac fermions,[34] ideal optical[35] and electrochemical properties,[36-40] the catalysis capacity of $PtSe_2$ for HER and OER is not clear. Previous studies indicate defects in h-BN,[7] silicene,[8,9] phosphorene[14] and $MoS_2$[17-19] indeed enhance the catalysis behaviors. However, studies on the structure and catalysis capacity of defects, especially multiple ones, in $PtSe_2$ are still rare.[41-53]

In this paper, the structures, fractions and catalytic performance of various Se vacancies in defective PtSe$_2$ (d-PtSe$_2$) monolayers were investigated systematically. We indicate the point Se vacancies are quite stable once they are introduced in PtSe$_2$. The migration barrier is over 3 eV, therefore, Se vacancies are located steadily in their positions without diffusion. Besides, every Se vacancy indeed traps local charge, which benefits the catalysis. With consideration of the water solvation effect, our ab initio calculations predict Se-1V, 2V and 3V are all very good for HER catalysis. Most importantly, we firstly reveal Se-2V exhibits excellent OER catalysis, which is rare in Pt-based materials.[54,55] Beyond, the experimental measurements find defective PtSe$_2$, synthesized by using polycrystal Pt foil, has excellent performance in both HER and OER indeed. The good agreement between theoretical and experimental results show defective PtSe$_2$ is a truly fascinating bipolar catalyst. This research not only provides a basic foundation for theoretical and experimental synergetic studies to search good catalyst, but also promotes the PtSe$_2$ applications in water splitting.

## 2. Results and Discussion

### 2.1. Se Defects in Monolayer PtSe$_2$

Se defects will break the spatial symmetry of two surfaces of 1T sandwich structure PtSe$_2$. Unlike the widely studied Pt defects which induce a local magnetic moment,[41-44,47,52,53] Se defects improve the electronic conductivity.[46,48] Note that some different Se defects even exhibit nearly same scanning tunneling microscope (STM) images, making the real structures of Se defects are very hard to be identified in experiment while Pt defects are relatively easy to identify.[45] Therefore, the prediction of STM images corresponding to different Se defects by first-principles calculations is important. For each point defect, there are two simulated STM images observing from the two surfaces (called top and bottom in Figure 1a-e).

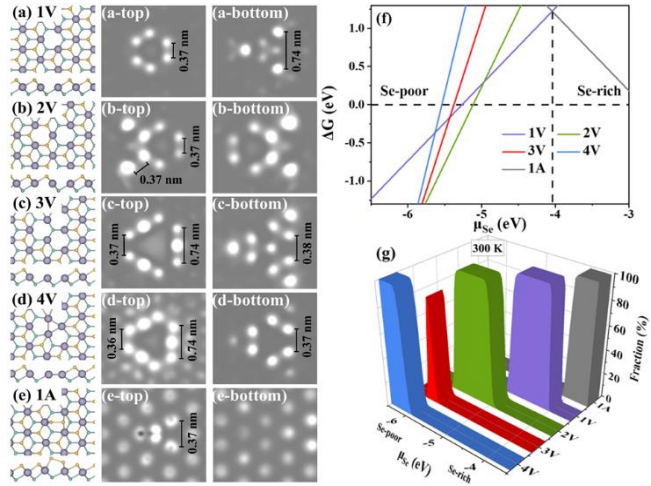

**Figure 1.** Top and lateral views of different Se defects in PtSe$_2$ monolayer: (a) Se-1V, (b) Se-2V, (c) Se-3V, (d) Se-4V and (e) Se-1A. Yellow and green balls represent Se atoms on the top and at the bottom, respectively. Purple balls represent Pt atoms. The simulated STM images obtained from top and bottom surfaces were calculated at −0.5 V bias. (f) Phase diagram and (g) Population of defective PtSe$_2$ monolayers under different μ$_{Se}$.

First, it is very difficult to connect STM image to real atomic structure of Se vacancy. For example, Se single vacancy from top surface (Se-1V) possesses hexagonal shape, while it is a triangle pattern from bottom surface (Figure 1a-top, bottom). For Se-2V, the STM images are both one and a half hexagonal patterns (Figure 2b-top, bottom). Another double vacancy named Se-2V' predicted by our previous work,[45] equivalent to two Se-1V side by side, was calculated as well, whose simulated STM images (Figure S2) can be regarded as two Se-1V ones overlap with each other. We also gave the result of Se-3V from bottom surface which shows fractal triangle with ten spots. This is quite different from STM images of defects in graphene,[56,57] which are much easier to identify due to its planar monolayer structure.

Besides, the STM images of Se-1V from top surface (Figure 1a-top) and Se-4V from bottom surface (Figure 4d-bottom) are quite similar. Se-1V is formed by losing one top Se atom without obvious deformation. From top surface, the STM image is hexagonal shape consisted of six bright spots with 0.37 nm distance. The STM image of Se-4V from bottom surface nearly resembles that of Se-1V. A slight difference is there are three additional dim spots around the bright hexagonal spots.

Another similar STM images occurs on top surface of Se-3V (Figure 1c-top) and Se-4V (Figure 1d-top). They both exhibit truncated regular triangle with nine bright spots. The edge lengths are both about 0.74 nm, and the middle spots in the edge are brighter than the neighbors. The contrast of brightest spots to other spots and substrates in STM images of Se-3V is sharper that of Se-4V, which may slightly distinguish them.

Surface adatom is also common defect in 2D materials. Interestingly, Se-1A will pull one neighboring Se atom, forming a bulge bridge structure on the top surface. This is unique comparing with graphene and silicene. The STM image of Se-1A from top surface (Figure 1e-top) just looks like double vacancy (5|8|5) defect in graphene and silicene.[8,56,57] While from bottom surface, Se-1A just like having no defects (Figure 1e-bottom).

Previous studies on 2D materials have indicated the population of various vacancies are dependent on their free energy change.[58] As $PtSe_2$ is usually produced by direct selenization of Pt foils (Details are given in Table S9),[31,59-68] the chemical potential of Se vapor ($\mu_{Se}$) played a significant role. Here, the free energy change of Se vacancy is related to $\mu_{Se}$ by the formula:

$$\Delta G = E_{defect} - NE_{PtSe_2} + n\mu_{Se} \qquad (1)$$

where $E_{defect}$ is the total energy of d-$PtSe_2$ models, and $E_{PtSe_2}$ is the energy per ($PtSe_2$) unit in perfect $PtSe_2$ monolayer, respectively. The free energy of our explored defects as a function of $\mu_{Se}$ is plotted in Figure 1f while a more detailed diagram including Pt defects is available in Figure S3. During growth of $PtSe_2$, the value of $\mu_{Se}$ must be restricted to obtain better $PtSe_2$. Perfect $PtSe_2$ is the lowest-energy structure for $\mu_{Se}$ from -5.13 to -2.82 eV. At this range, the most probable defects are Se-1V, 2V and 1A, because they have relative lower free energy. The cross point of Se-1V and 1A indicates the balanced value of $\mu_{Se}$, whose left is Se-poor while right is Se-rich region.

We can quantitatively evaluate the population (Figure 1g) of various defects in PtSe$_2$. From the thermodynamic point, the population of defect is proportional to exp(-ΔG/k$_B$T), where k$_B$ and T is the Boltzmann constant and temperature, respectively. Therefore, the related fraction can be estimated by $c = N_i/\sum N_i$. At Se-rich region, Se-1A is dominating at μ$_{Se}$ ∈ (-2.82, -1.71) eV. Pt defects are calculated to be popular at very Se-rich level (μ$_{Se}$ > -1.71 eV, Figure S3). Thus, in the growth of PtSe$_2$ monolayer, the Se amount must be avoid too rich.

At Se is slightly poor but ΔG > 0 region, Se-1V is dominating at μ$_{Se}$ ∈ (-4.96, -4.04) eV. The highest population of Se-1V can reach up to 99.9 % at this region. Then Se-2V is the dominating defect at μ$_{Se}$ from -5.13 to -4.96 eV. Further reduce μ$_{Se}$ then ΔG will be negative, which means defective PtSe$_2$ is more stable than pristine. At this region, Se-2V is the most probable defect at very large μ$_{Se}$ range from -5.89 to -5.13 eV with highest fraction of 99.9 %. Clearly, Se-1V and Se-2V are the most probable defects according to Figure 1g, their relative fractions can be tuned by the amount of Se. Large defects like Se-3V (μ$_{Se}$ ∈ (-6.02, -5.89) eV) and Se-4V (μ$_{Se}$ < -6.02 eV) can only exist in Se very poor region.

**2.2. Kinetics Behavior and Stability of Se Vacancy**

The migration of defects on the surface of 2D-materials plays an essential role in their performance, such as stability, reliability and recyclability. According to previous work,[45] the migration of single vacancy in PtSe$_2$ including Se-1V and Pt-1V both shows high kinetic diffusion barriers range from 2 ~ 3 eV. But in practice, vacancies will not appear alone. It is necessary to consider whether there are multiple vacancy defects existing at the same time. We investigated a typical case of two Se single vacancy gather into a larger one. As shown in Figure 2, Se vacancies on the same side and different side of PtSe$_2$ monolayer were investigated, both of them exhibit extremely huge barriers. We had considered more complicated cases (Figure S4) and get the same results which depict a huge diffusion barrier up to 4.20 eV.

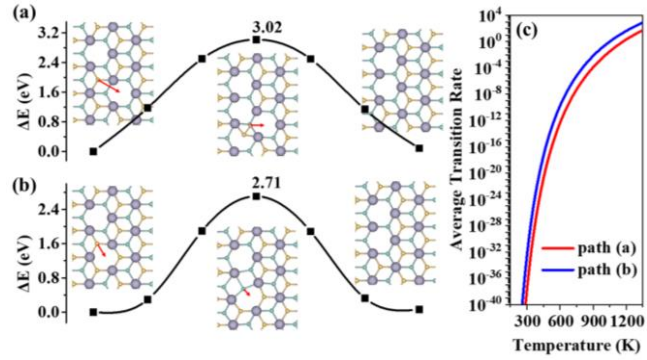

**Figure 2.** (a, b) Possible pathways where there are two Se single vacancies. (c) Average transition rate under different temperature.

From the point of atomic vibration, the average transition rate can be evauated by the Arrhenius form ν*exp(-E/$k_B$T), where the prefactor ν is approximately equal to $10^{13}$ Hz under 300 K, which is vibration frequency in solid. E denotes the barrier, $k_B$ and T is the Boltzmann constant and temperature, respectively. Under room tempearture, the two typical pathways only possess extremely low rate about $10^{-40}$ Hz. This means Se vacancy can hardly migrate on $PtSe_2$ monolayer. Even if the temperature is heated to 1200 K, the value of rate just reaches up to $10^0$ Hz, which is still unconspicuous.

Thus, once a defect in $PtSe_2$ is formed, it is hard to migrate. That means it is highly stable against thermal perturbance without gathering into larger defect or being recovered, which is consistent with previous experiment.[69] From this point of view, if we want perfect $PtSe_2$, we must carefully optimize the growth technology from the beginning. But if we want to utilize the defect, the permanent defects can be designedly introduced during the growth or in postprocessing irradiation processs.

### 2.3. Defective Level and Electron Trapping of Defective $PtSe_2$

$PtSe_2$ is semiconductor with an indirect band gap of 2.10 eV using GW method[31] and 1.40 eV using PBE method.[45] For Se-1V (Figure 3a), there are three defective levels emerging near the Fermi level. The real space charge density of lowest unoccupied molecular orbital (LUMO) and the highest occupied molecular orbital (HOMO) indicate these flat bands are contributed by electron localization near the defect. In order to distinguish the source of trapping electrons, partial density of state (PDOS, Figure S5) of atoms around the vacancy were calculated carefully. For instance, $d_{yz}$ orbital of three nearest Pt atoms and $p_x$ orbital of three next nearest bottom Se atoms are trapped by the defective levels of Se-1V.

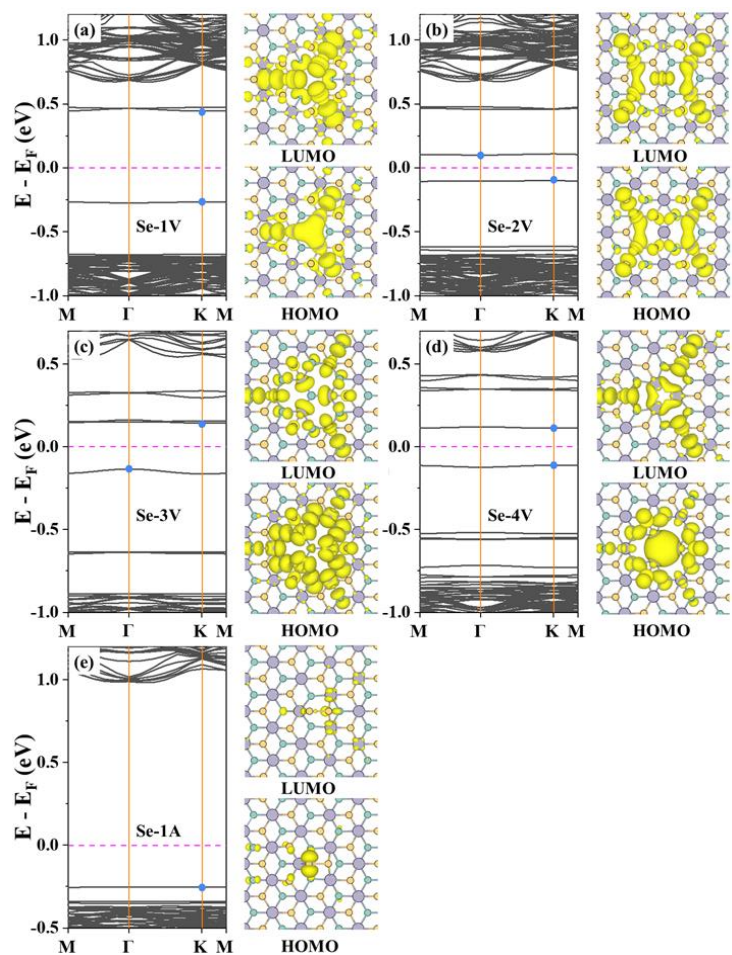

**Figure 3.** Band structures and decomposed charge images of various Se defects in PtSe$_2$ monolayer: (a) Se-1V, (b) Se-2V, (c) Se-3V, (d) Se-4V and (e) Se-1A (isosurface level = 0.001 |e|/bohr$^3$). Blue spots denote the lowest points of LUMO and highest points of HOMO.

For Se-2V, localized electrons come from $d_{xz}$ orbital of Pt atoms and $p_z$ orbital of Se atoms near the vacancy. The only difference between its LUMO and HOMO is that the $d_{x^2}$ and $d_{xy}$ orbitals of center Pt atom only contributed to the LUMO. Besides, Se-2V exhibits a narrowest gap among all Se vacancy defective configurations. In this case, the excitation of electrons is more likely to occur in Se-2V,[70] which benefits its applications in photo-recombination and optical excitation.

Se-2V', together with large Se-3V (Figure 3c) and Se-4V (Figure 3d) have similar flat defective bands, and their contributed charge can be found in Figure S5. Se vacancies leads to electron localization near the defective region and induces the appearance of defective levels, which have trapping effect on electrons mostly from p orbital of Se atom and d orbital of Pt atom. Besides, Se defects will also enhance the conductivity of $PtSe_2$,[46,48] improving the catalytic performance.

Different from Se vacancy, only one defective level will be introduced under Fermi level in Se-1A (Figure 3e). Electron of its LUMO distribute almost unanimously the same with perfect $PtSe_2$,[45] while numerous electron from Se atoms around the defect contribute a lot to the HOMO. Weaker localization of electron makes Se-1A have no occasion for catalyst comparing with vacancy defects.

Clearly, our DFT calculations indicate the Se defects generally introduce flat defetive levels in their band structures, which will be the charge trapping centers. Such trapping centers may influence the eletronic transport behaviors, which should be taken care in the application of $PtSe_2$. Most of important, such defective levels with trapping charge may be highly efficient active catalyst sites.

**2.4. Bipolar Catalysis of Defective $PtSe_2$ for HER and OER**

Pt is an ideal catalyst for HER and its (111) surface of bulk phase has been widely applied.[30] However, in addition to the surface Pt atoms, deeper Pt atoms did not participate in the catalytic reaction, resulting in a great waste. 2D $PtSe_2$ has two large surfaces but only one-layer of Pt, is ideal platform for catalysis.

We then explored the HER performance of PtSe₂ under alkaline environment using overvoltage, which is free energy changing divide by *e*, the electric charge of a single electron (details were given in the computational section)

$$\Delta G = E_a + \Delta E_{ZPE} - T\Delta S \quad (2)$$

$$E_{a(H*)} = E_{H*} - E_{*} - 1/2 E_{H^2} \quad (3)$$

where $E_a$ is the absorption energy of hydrogen, $\Delta E_{ZPE}$ represents the change of zero vibration energy, T and $\Delta S$ are temperature and entropy change. $E_{H*}$, $E_{*}$ and $E_{H^2}$ is energy of d-PtSe₂ adsorbed hydrogen, d-PtSe₂ and hydrogen, respectively. We detailly listed data of $E_{H^2}$, $\Delta E_{ZPE}$, T$\Delta S$ and $E_a$ in Table S2-4. Besides, the water solvation effect was taken into account due to interaction of water molecules with H-ending adsorption species. Detailed methodology was given in the computational section. The closer $\Delta G$ to zero, the better HER catalytic performance is.

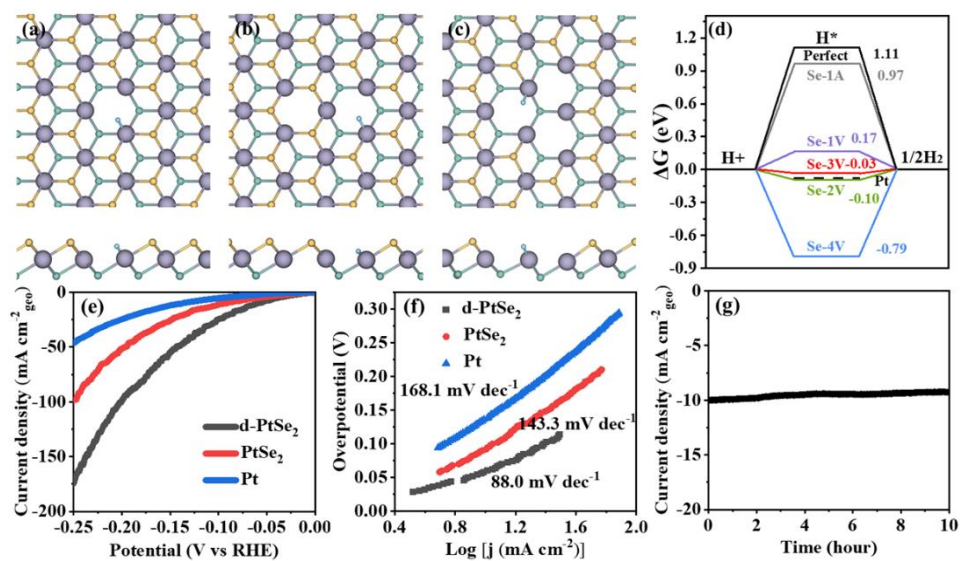

**Figure 4.** Top and lateral views of three typical defective PtSe₂ monolayers for HER: (a) Se-1V, (b) Se-2V and (c) Se-3V. Small blue balls represent H atoms. (d) Free energy during HER of different Se vacancy configurations. (e) LSV curve of d-PtSe₂, PtSe₂ and Pt in 1M KOH at scan rate of 5mV s⁻¹. (f) Tafel slope of d-PtSe₂, PtSe₂ and Pt. (g) Time-dependent current density curves of d-PtSe₂ at typical potential.

In perfect PtSe$_2$, hydrogen ion prefers to be adsorbed between Pt and Se atoms (Figure S6a). However, the perfect PtSe$_2$, whose ΔG is 1.11 eV, is not good at for HER (Figure 4d). Even if the water solvation effect is considered, the ΔG slightly change to 1.13 eV. We found the water solvation effect caused an about 0.3 ~ 0.4 eV upshift in total energy of d-PtSe$_2$ configurations. H$_2$ energy (Table S2), ZPE and vibration entropy (Table S3) are basically unaffected. Thus, the upshift energy is offset by subtraction during calculation of adsorption energy in HER. The water solvation effect has very little impact on ΔG of HER (about 0.01 eV, Table S4) and we will ignore it during HER discussion.

As we mentioned in section 2.3, Se vacancies always traps remarable local electron, which may enhance the catalyst capacity. We found Se-1V reduces the free energy of HER significantly from 1.11 eV to 0.17 eV, the hydrogen ion is adsorbed by one of the three equivalent Pt atoms near single vacancy. There are two adsorption sites in Se-2V. One is the Pt atom locates at the center of double vacancy which bonds to four neighbor Se atoms. The other is the four equivalent Pt atoms around the DV. Interesingly, the later sites are very good at HER (Figure 4b) with a small ΔG of -0.10 eV, which is comparing with -0.08 eV of Pt(111) already.[30] Se-3V possesses two types of adsorption sites as well, which are the three nearest neighbor Pt atoms and three next nearest neighbor Pt atoms around the vacancy. Hydrogen ion is more likely to be adsorbed on one of the three nearest neighbor Pt atoms. Among various defective PtSe$_2$ monolayers, Se-3V exhibits the lowest ΔG of -0.03 eV, less than half of Pt(111) surface, indicates its even better catalytic performance. We regarded Se-1V, Se-2V and Se-3V as ideal catalytic structures for HER. As shown in Figure 4d, free energies of these defective monolayers are equal to or even better than that of Pt.

Different from other defective PtSe$_2$ configurations, hydrogen ion is adsorbed on the bridge site of two nearest neighbor Pt atoms in Se-4V. The excessively stronger adsorption of hydrogen ion makes the overvoltage required for activation higher. Pt-1V and Pt-2V have weaker catalytic performance as well, in which hydrogen ion is strongly adsorbed on Se atom.

Se-1A is not suitable for catalysis as well due to the difficulty to adsorb hydrogen ion, which is quite similar to perfect PtSe$_2$, thus, huge free energy of these two configurations is nothing surprising. Models and data of above defects during HER were given in Figure S6 and Table S4, respectively.

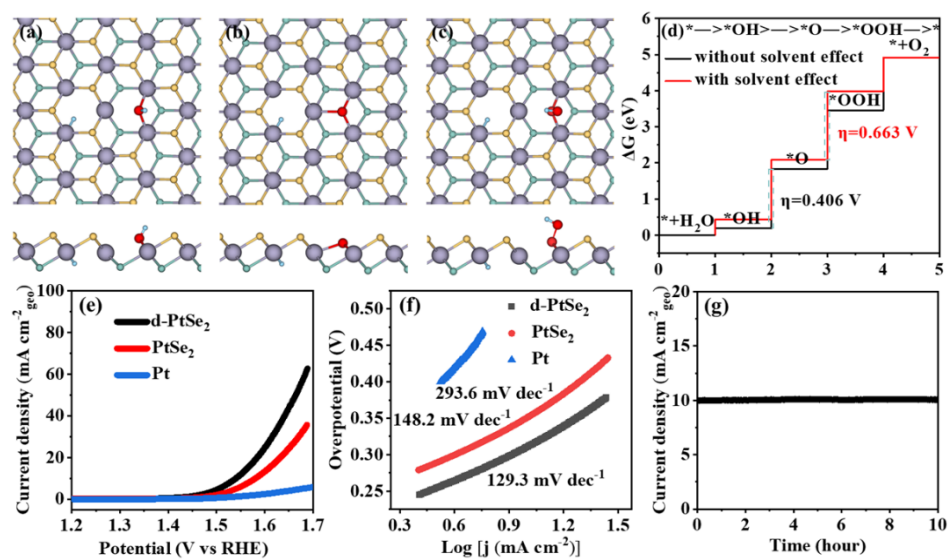

**Figure 5.** Top and lateral views of Se-2V for OER processes: (a) Se-2V+*OH, (b) Se-2V +*O, (c) Se-2V +*OOH. Blue and red balls represent H and O atoms. (d) Free energy of hydrogenated Se-2V during OER without/with the solvent effect. (e) LSV curve of d-PtSe$_2$, PtSe$_2$ and Pt in 1M KOH at scan rate of 5mV s$^{-1}$. (f) Tafel slope of d-PtSe$_2$, PtSe$_2$ and Pt. (g) Time-dependent current density curves of d-PtSe$_2$ at typical potential.

In the experimental study, we synthesized PtSe$_2$ by carefully investigating the previous CVD process of PtSe$_2$ referring the reports[31,59-68] in Table S9. We found that the Se content, reaction temperature (400 ~ 500 °C) and gas flow (H$_2$/Ar) affected the sample quality of PtSe$_2$ remarkably. Se powder in high purity was placed at the upstream position of the tube to supply a continuous Se source and introduced defects by tuning the deposition time. Then we chose 400 ~ 500 °C as growth temperature of d-PtSe$_2$, where Se powder is evaporated in small amounts (Detailed synthesis process was given in the experimental section). Various Physical characterizations of PtSe$_2$ including X-ray diffraction (XRD, Figure S14), X-ray photoelectron spectroscopy (XPS, Figure S16), transmission electron microscopy (TEM, Figure S18) and electron paramagnetic resonance (EPR, Figure S19) spectroscopy were employed to verify the defects have been successfully introduced in PtSe$_2$ sample.

Electrochemical performance of PtSe$_2$ was conducted using a three-electrode configuration through linear scan voltammogram (LSV) in 1M KOH. For comparison, Pt was measured under the same condition. Detailed measurement process was given in the expreimental section. The LSV curve was exhibited in Figure 4e, d-PtSe$_2$ achieves a very low overvoltage of 59 mV at the current density of 10 mA cm$^{-2}$, which is even better than those of PtSe$_2$ and Pt foil. The hydrogen evolution kinetics of d-PtSe$_2$ was calculated through corresponding Tafel plots. The derived Tafel slope of 88 mV dec$^{-1}$ for d-PtSe$_2$ indicated the Volmer-Heyrovsky mechanism as the HER pathway for PtSe$_2$. Moreover, the long-term stability is a pivot criterion for electrocatalyst as shown in Figure 4g, in which d-PtSe$_2$ exhibits excellent stability with the negligible degradation in 10 h.

It is well known that Pt slab is not good for OER. We also found perfect PtSe$_2$ monolayer has an extremely high barrier for OER. The models are given in Figure S7 and data could be found in Table S5,6. Catalytic performance of d-PtSe$_2$ monolayers for OER under adsorbates evolution mechanism (AEM,[71] given in the computational section) were investigated as well. Detailed results are available in Table S5,6. In Se-1V, 2V and 3V with good HER catalytic performance, the adsorption of -OH, -O and -OOH were carefully tested. Oxygen atoms tend to fill the Se vacancy due to O and Se are both elements of Group VI (Figure S9-11). Since oxygen possesses stronger electronegativity, the length of Pt-O bond is shorter than Pt-Se bond. According to Table S6, overvoltage of most Se defects is higher than 1 V except Se-2V performs a lower overvoltage of 0.854 V at U$_{RHE}$ = 1.23 V, where U$_{RHE}$ is the overvoltage provided by water. The biggest step locates between -O and -OOH in these structures.

More than our expectation, introduction of H atom on the other side of Se-2V will enhance the adsorption of -O, which shortens the step between -O and -OOH as shown in Figure 5. The overvoltage drops down to only 0.406 V. This is a rare good value for OER in Pt-based materials except for newly proposed $Pt_5Se_4$.[54,55] We have compared the catalytic performance of our d-$PtSe_2$ and $Pt_5Se_4$ and found they are very close. Models and data are available in Figure S1 and Table S1,4-6, respectively. Owing to the interaction between water molecules with H-ending adsorption species like -OH and -OOH, water solvation effect is further considered. This correction will cause an about 0.3 ~ 0.4 eV upshift in total energy of d-$PtSe_2$ configurations just like HER. $H_2$ energy, ZPE and vibration entropy are still basically unaffected. However, $H_2O$ energy has a 0.3 eV downshift, which leads to the general increase of -OH, -O and -OOH adsorption energies. After water solvation effect is considered, the overvoltage of hydrogenated Se-2V will rise slightly to 0.663 V, which is still an ideal value in practical. In other words, an improved catalysis for OER gets probable when Se-2V is during HER. A water molecule dissociation with barrier of about 1.4 eV near Se-2V was given in Figure S12, which shows multiple active sites and parallel process of Se-2V. Referring to previous study of 1.97 eV,[72] such a water dissociation barrier can ensure the HER and OER occurs under alkaline environment. Se-2V have multiple active sites for water dissociation, thus it could catalyst in a parallel mode. The same rule is also applicable to Se-3V (Figure S11), that is, hydrogenation will enhance the OER performance[73] of d-$PtSe_2$.

The OER performance of d-$PtSe_2$, $PtSe_2$ and Pt foil was also evaluated in oxygen-saturated 1M KOH. The d-$PtSe_2$ possesses an excellent OER activity which achieves the current density of 10 mA $cm^{-2}$ at 1.54 V, which is much higher than those of $PtSe_2$ and Pt foil. Tafel slope of d-$PtSe_2$ was evaluated to be 129.3 mV $dec^{-1}$, lowest among the three samples. Moreover, d-$PtSe_2$ preserves the OER activity at 10 mA $cm^{-2}$ over 10 h, indicating its fascinating OER stability. XRD and XPS were performed again to reveal the characteristics of catalyst itself. In XRD (Figure S15), the position of peaks did not change, but the number of peaks significantly became less, indicates the decrease degree of crystallization, which may be caused by surface oxidation after OER.[74,75] In XPS (Figure S17), the peak locates at 58.4 eV newly appeared corresponds to bonding between Se and O. Referring to curves of O 1s, there are two peaks locate at 530.4 eV and 531.4 eV, which correspond to lattice and surface hydroxyls during OER.[76,77] We also adopted the double-layer capacitance extracted from the cyclic voltammetry (CV) measurement (Figure S20) and the electrochemical impedance spectroscopy (Figure S21) for comparison purposes. All the results are in good agreement with the theoretical prediction and demonstrate convincingly that d-$PtSe_2$ performs a truly better bipolar catalysis than Pt.

Catalytic activity of $PtSe_2$ monolayer can be greatly enhanced by localized electron near the Se vacancy. Comparing with traditional catalyzer such as representative noble metal Pt, d-$PtSe_2$ maintains efficiency and economy. It is feasible to practical applications due to the stability of d-$PtSe_2$. In general, specific Se vacancy defects can be introduced to achieve bipolar catalysis of $PtSe_2$ monolayer. As for other defective 2D Pt-based materials such as $PtS_2$ and $PtTe_2$,[78-83] the similar mechanism exists as well (Figure S13, Table S7,8). Our results provide a new way for bipolar catalysis of HER and OER for Pt-based materials. Defect regulation can significantly enhance the catalytic performance of materials, brings them broader prospect in future applications.

## 3. Conclusion

In summary, we systematically investigated Se vacancy defects in $PtSe_2$ monolayer by first-principle calculations and experimental verification. We gave detailed STM images of various d-$PtSe_2$ monolayers, which is very helpful to distinguish those defects with great similarity. Population of typical defects under different chemical potential of Se vapor were quantitatively evaluated. Also we declared the stability of d-$PtSe_2$ and feasibility to regulate defective configuration due to the investigation on thermodynamic properties and kinetic behavior of Se atom. Introduction of Se vacancy will induce the localization of electron near the defect, which greatly enhances the catalytic performance. For instance, we highlighted the ideal bipolar catalysis of Se-2V for HER and OER, which promises regulated $PtSe_2$ monolayer competitive for electrocatalytic applications. Our work fills the blank of understanding about

d-PtSe$_2$ and provides a successful case which combines theoretical prediction and experimental verification together for material preparation and application

## 4. Computational Section

*Computational Method*: We employed the Vienna ab initio simulation (VASP)[84,85] package for first-principle calculations. The ion-electron interaction was treated with the projected-augmented wave (PAW) method[86] using Perdew-Burke-Ernzerhof (PBE) pseudopotentials.[87] We adopted 8×8×1 supercell of PtSe$_2$ under gamma point for calculating, which ensures enough spacing between periodic vacancies. To compare the catalytic performance, we also carried out calculations on 2×2×1 and 5×5×1 supercell of Pt$_5$Se$_4$.[54,55] More information of our models could be found in Figure S1 and Table S2. Defective PtSe$_2$ will probably encourage a magnetic moment like MoS$_2$, therefore spin polarization effect was considered. We also considered the water solvation effect of H-ending adsorption species, such as -H, -OH and -OOH by using VASPsol.[88] During structure optimization, a competent kinetic energy cutoff of 400 eV was utilized. The total energy convergence criterion was set to $10^{-4}$ eV while the force criteria for structure optimization was set to be 0.02 eV/Å. During scanning tunneling microscope image simulation using the Tersoff–Hamann approximation,[89] a −0.5 V bias was utilized. As for migration calculations we used the climbing image nudged elastic band (CI-NEB) method,[90,91] with 5 images adopted to find transition phases. Based on previous work,[92] we added van der Waals correction due to its essential role in PtSe$_2$. All the accuracy of our numerical procedure had been carefully tested.

*Reaction Mechanism of HER*: The process of HER can be described as following relations under alkaline environment

$$H_2O + * + e^- \rightarrow {}^*H + OH^-  \quad (4)$$

$$H_2O + {}^*H + e^- \rightarrow H_2 + OH^- + * \quad (5)$$

Free energy could be described by the equation

$$\Delta G = E_a + \Delta E_{ZPE} - T\Delta S \tag{6}$$

Overvoltage during the reaction is free energy changing divide by $e$, the electric charge of a single electron

*Reaction Mechanism of OER*: The process of OER can be described as following relations under alkaline environment

$$* + OH^- \rightarrow {^*OH} + e^- \tag{7}$$

$$^*OH + OH^- \rightarrow {^*O} + H_2O + e^- \tag{8}$$

$$^*O + OH^- \rightarrow {^*OOH} + e^- \tag{9}$$

$$^*OOH + OH^- \rightarrow O_2 + H_2O + e^- + * \tag{10}$$

free energy of different absorption could be obtained by the equation

$$E_{a(^*OH)} = E_{^*OH} - E_* - \left(E_{H_2O} - \frac{1}{2}E_{H_2}\right) \tag{11}$$

$$E_{a(^*O)} = E_{^*O} - E_* - \left(E_{H_2O} - E_{H_2}\right) \tag{12}$$

$$E_{a(^*OOH)} = E_{^*OOH} - E_* - \left(2E_{H_2O} - \frac{3}{2}E_{H_2}\right) \tag{13}$$

next the free energy corresponding to each step could be calculated

$$\Delta G_1 = \Delta G_{^*OH} \tag{14}$$

$$\Delta G_2 = \Delta G_{^*O} - \Delta G_{^*OH} \tag{15}$$

$$\Delta G_3 = \Delta G_{^*OOH} - \Delta G_{^*O} \tag{16}$$

$$\Delta G_4 = 4.92 - \Delta G_{^*OOH} \tag{17}$$

$$G^{OER} = \max\{\Delta G_1, \Delta G_2, \Delta G_3, \Delta G_4\} \tag{18}$$

$$\eta^t = G^{OER}/e - 1.23\ V \tag{19}$$

where $G^{OER}$ represents the free energy of OER while $\eta^t$ is the required overvoltage for oxygen evolution.

## 5. Experimental Section

*Materials and Synthesis*: PtSe$_2$ was synthesized using a chemical vapor deposition (CVD) process. It showed that the Se content, reaction temperature (400 ~ 500 °C) and gas flow (H$_2$/Ar) affected obviously the sample quality of PtSe$_2$ according to previous work (Table S9). Thus, we placed Se powder (Aladdin, 200mg, purity 99.9%) at the upstream position of the tube to supply a continuous Se source and introduce defects by tuning the deposition time. We adopted Pt foil (Japan, 0.1 mm, purity 99.99%) as substrate, which was pretreated with diluted HCl, and then cleaned by deionized water and ethanol for several times. After the surface was dry it was put in the down-stream of the tube. As for reaction temperature, a series of experiments were carried out with a gradient of 100 °C. We found 400 ~ 500 °C is the ideal temperature, where Se powder is evaporated in small amounts. Finally, the furnace was gradually heated from room temperature to 500 °C at a rate of 5 °C/min under 40 sccm Ar gas flow. The CVD system was maintained at 500 °C for 2h and then cooled down to room temperature naturally. As-obtained PtSe$_2$ were irradiated by RF plasma under an Ar flow (RF power, 50~150 W).

*Characterization*: Powder X-ray diffraction (XRD) patterns were tested by X-ray diffractometer (Japan Rigaaku Rotalex) by Cu K$_\alpha$ radiation ($\lambda$ = 1.5418 Å). TEM images were performed on FEI TF30. The electron paramagnetic resonance (EPR) tests were performed through Bruker 500 spectrometer (Bruker E500). X-ray photoelectron spectroscopy (XPS, ESCALAB 250) pattern was applied to explore the elements composition and valence states of materials.

*Electrochemical Measurements*: The electrochemical performance of the samples was investigated using a CHI 760 electrochemical workstation (Chenhua) through a typical three-electrode configuration in N$_2$ or O$_2$-saturated 1M KOH electrolyte for HER and OER. The PtSe$_2$, graphite rod and Hg/HgO were used as the working electrode, counter electrode and reference electrode, respectively. All applied potentials were converted with respect to reversible hydrogen electrode (RHE), $E_{RHE} = E_{Hg/HgO} + 0.059$ pH $+ 0.098$ V. Polarization curves were correct with iR compensation. EIS was performed within the frequency range from 100 kHz to 0.1 Hz.

**Acknowledgements**

This work is supported by the National Natural Science Foundation of China (Grant No. 12074053, 91961204, 12004064) and by XinLiaoYingCai Project of Liaoning province, China (XLYC1907163). H. L. thanks the start-up funding (DUT20RC(3)026). We also acknowledge Computers supporting from Shanghai Supercomputer Center, DUT supercomputing center, and Tianhe supercomputer of Tianjin center. Yuan CHANG and Panlong Zhai contributed equally to this work.